\documentclass[english]{article}
\usepackage{graphicx}
\usepackage{amssymb}

\makeatletter

\providecommand{\tabularnewline}{\\}

\usepackage{amsmath,spconf}
\makeatother
\begin{document}
\name{Jean-Marc Valin, Jean Rouat and Fran\c cois Michaud\thanks{
This research is funded by the Natural Sciences and Engineering Research Council, the Canada Research Chair Program and the Canadian Foundation for Innovation.\break
\copyright 2004 IEEE.  Personal use of this material is permitted. Permission from IEEE must be obtained for all other uses, in any current or future media, including reprinting/republishing this material for advertising or promotional purposes, creating new collective works, for resale or redistribution to servers or lists, or reuse of any copyrighted component of this work in other works.
}}\address{Department of Electrical Engineering and Computer Engineering, Universit\'e de Sherbrooke\\\texttt{jmvalin@jmvalin.ca}\vspace{-0.05cm}}

\title{Microphone array post-filter for separation of simultaneous non-stationary
sources}

\maketitle
\begin{abstract}
Microphone array post-filters have demonstrated their ability to greatly
reduce noise at the output of a beamformer. However, current techniques
only consider a single source of interest, most of the time assuming
stationary background noise. We propose a microphone array post-filter
that enhances the signals produced by the separation of simultaneous
sources using common source separation algorithms. Our method is based
on a loudness-domain optimal spectral estimator and on the assumption
that the noise can be described as the sum of a stationary component
and of a transient component that is due to leakage between the channels
of the initial source separation algorithm. The system is evaluated
in the context of mobile robotics and is shown to produce better results
than current post-filtering techniques, greatly reducing interference
while causing little distortion to the signal of interest, even at
very low SNR.
\end{abstract}

\section{Introduction}

Mobile robots with abilities to talk and listen should be able to
discriminate and separate simultaneous sound sources while moving.
For example, in the context of the cocktail party effect, the algorithms
have to be robust and should allow the separation of simultaneous
voices. 

In the present work we first perform a crude linear separation of
the sources and then use the proposed post-filter to further enhance
the signals and suppress the contribution of the perturbating sources.
Our post-filter is inspired by the original work of Cohen \cite{CohenArray2002}
who proposes a post-filter designed for a beamformer with one source
of interest in the presence of stationary background and transient
noises. In the present work we extend the principle to multiple localized
sources of interest.

We assume that both the signal of interest and the interferences may
be present at the same time and for the same frequency bin. The novelty
of our approach resides in the fact that, for each source of interest,
we decompose the noise estimate into a stationary and a transient
component assumed due to leakage between channels occuring during
the initial separation stage.

For each output channel of the linear source separator, we adaptively
estimate the interference parameters (variance and SNR) and use them
1) to compute the probability of targeted speech presence 2) in the
suppression rule when both speech and interference are present.

We also propose the use of a Minimum Mean Square Estimation (MMSE)
of the loudness -- instead of the common log amplitude estimation
-- yielding a more efficient cleaning of the signal when targeted
speech is not present in the channel of interest.

Section \ref{sec:System-overview} gives an overview of the system
and Section \ref{sec:Loudness-domain-spectral-attenuation} describes
the proposed post-filter. Results and discussion are then presented
in Section \ref{sec:Results} with the conclusion in Section \ref{sec:Discussion}.

\section{System overview}

\label{sec:System-overview}

The source separation system discussed here is composed of two subsystems:
1) a linear source separation (LSS) algorithm and 2) the proposed
post-filter (Fig. \ref{cap:Overview-System}). By \emph{linear separation
algorithm}, we mean any separation algorithm for which a channel output
is the result of a linear transformation of the microphone signals.
Most Blind Source Separation (BSS) algorithms fall in this category,
as well as distortion-less beamformers and Geometric Source Separation
(GSS) techniques \cite{Parra2002Geometric}. 

The Linear Source Separation system used for our experiments is inspired
from the second constrained (C2) meth\-od in \cite{Parra2002Geometric}
and comprises

\begin{enumerate}
\item The localization algorithm such as the one described in \cite{ValinIROS2003}
-- It is based on the Time Delay of Arrival (TDOA) estimation;
\newpage
\item The estimated mixing matrix -- Assuming unity gain for all microphones,
while the phases are computed from the localization algorithm;
\item The pseudo-inverse of the estimated mixing matrix.
\end{enumerate}
In the design of the proposed system, we already take into account
that the final application is mobile robotics. As a consequence, our
implementation of the LSS system does not include any iterative algorithm
-- by the time convergence is reached, the robot (or one source) has
already moved. We are aware that the LSS algorithm is far from perfect
(hence the need for a post-filter) because of localisation accuracy,
reverberation and imperfect microphones (non-iden\-tical response).
We design the post-filter in such a way that any source separation
algorithm (including blind algorithms that do not require localization
of the sources) can be used.

\section{Loudness-domain spectral attenuation}

\label{sec:Loudness-domain-spectral-attenuation}%
\begin{figure}
\includegraphics[%
  width=1.0\columnwidth,
  keepaspectratio]{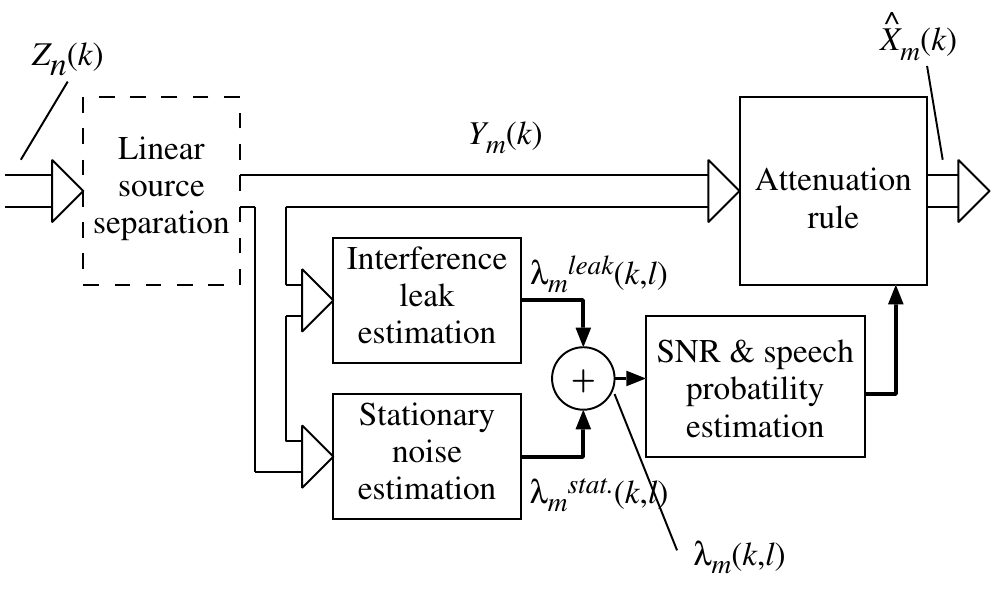}

\caption{Overview of the complete separation system. }

$Z_{n}(k,l),n=0\ldots N-1$: Microphone inputs, $Y_{m}(k,l),\: m=0\ldots M-1$:
Inputs to the post-filter, $\hat{X}_{m}(k,l)=G_{m}(k,l)Y_{m}(k,l),\: m=0\ldots M-1$:
Post-filter outputs.\label{cap:Overview-System}
\end{figure}

We derive a frequency-domain post-filter that is based on the optimal
estimator originally proposed by Ephraim and Malah \cite{EphraimMalah1984,EphraimMalah1985}.
The novelty of our approach resides in the fact that, for a given
channel output of the LSS, the transient components of the corrupting
sour\-ces is assumed to be due to \emph{leakage} from the other channels
during the LSS process. Furthermore, for a given channel, the stationary
and the transient components are combined into a single noise estimator
used for noise suppression, as shown in Figure \ref{cap:Overview-System}. 

For this post-filter, we consider that all interferences (except the
background noise) are localized (detected) sour\-ces and we assume
that the leakage between channels is constant. This leakage is due
to reverberation, localization error, differences in microphone frequency
responses, near-field effects, etc.

The next subsection describes the estimation of noise variances that
are used to compute the weighting function $G_{m}$ by which the outputs
$Y_{m}$ of the LSS is multiplied to generate a cleaned signal which
spectrum is denoted $\hat{X}_{m}$.

\subsection{Noise estimation}

The noise variance estimation $\lambda_{m}(k,l)$ is expressed as:\begin{equation}
\lambda_{m}(k,l)=\lambda_{m}^{stat.}(k,l)+\lambda_{m}^{leak}(k,l)\label{eq:noise_estim}\end{equation}
where $\lambda_{m}^{stat.}(k,l)$ is the estimate of the stationary
component of the noise for source $m$, at frame $l$, for the $k^{th}$
frequency component and $\lambda_{m}^{leak}(k,l)$ is the estimate
of source leakage.

We compute the stationary noise estimate $\lambda_{m}^{stat.}(k,l)$
using the Minima Controlled Recursive Average (MCRA) technique proposed
by Cohen \cite{CohenNonStat2001}. 

To estimate $\lambda_{m}^{leak}$ we assume that the interference
from other sources is reduced by a factor $\eta$ (typically $-10\:\mathrm{dB}\leq\eta\leq-5\:\mathrm{dB}$)
by the separation algorithm (LSS). The leakage  estimate is thus expressed
as:\begin{equation}
\lambda_{m}^{leak}(k,l)=\eta\sum_{i=0,i\neq m}^{M-1}S_{i}(k,l)\label{eq:leak_estim}\end{equation}
where $S_{m}(k,l)$ is the smoothed spectrum of the $m^{th}$ source,
$Y_{m}(k)$, and is recursively defined (with $\alpha_{s}=0.7$) as:\begin{equation}
S_{m}(k,l)=\alpha_{s}S_{m}(k,l-1)+(1-\alpha_{s})Y_{m}(k,l)\label{eq:smoothed_spectrum}\end{equation}

\subsection{Suppression rule in the presence of speech}

We now derive the suppression rule under $H_{1}$, the hypothesis
that speech is present. From here on, unless otherwise stated, the
$m$ index and the $l$ arguments are omitted for clarity and the
equations are given for each $m$ and for each $l$.

The proposed noise suppression rule is based on MMSE estimation of
the spectral amplitude in the loudness domain, $\left|X(k)\right|^{1/2}$.
The choice of the loudness domain over the spectral amplitude \cite{EphraimMalah1984}
or log-spectral amplitude \cite{EphraimMalah1985} is motivated by
better results obtained using this technique, mostly when dealing
with speech presence uncertainty (Section \ref{sec:Optimal-gain-modification}).

The loudness-domain amplitude estimator is defined by:\begin{equation}
\hat{A}(k)=\left(E\left[\left|X(k)\right|^{\alpha}\left|Y(k)\right.\right]\right)^{\frac{1}{\alpha}}=G_{H_{1}}(k)\left|Y(k)\right|\label{eq:amplitude_estim}\end{equation}
where $\alpha=1/2$ for the loudness domain and $G_{H_{1}}(k)$ is
the spectral gain assuming that speech is present. 

The spectral gain for arbitrary $\alpha$ is derived from Equation
13 in \cite{EphraimMalah1985}:\begin{equation}
G_{H_{1}}(k)=\frac{\sqrt{\upsilon(k)}}{\gamma(k)}\left[\Gamma\left(1+\frac{\alpha}{2}\right)M\left(-\frac{\alpha}{2};1;-\upsilon(k)\right)\right]^{\frac{1}{\alpha}}\label{eq:Gain_H1}\end{equation}
where $M(a;c;x)$ is the confluent hypergeometric function, $\gamma(k)\triangleq\left|Y(k)\right|^{2}/\lambda(k)$
and $\xi(k)\triangleq E\left[\left|X(k)\right|^{2}\right]/\lambda(k)$
are respectively the \emph{a posteriori} SNR and the \emph{a priori}
SNR. We also have $\upsilon(k)\triangleq\gamma(k)\xi(k)/\left(\xi(k)+1\right)$
\cite{EphraimMalah1984}.

The \emph{a priori} SNR $\xi(k)$ is estimated recursively as:\begin{equation}
\hat{\xi}(k,l)\!=\!\alpha_{p}G_{H_{1}}^{2}\!(k,l-1)\gamma(k,l-1)\!+\!(1-\alpha_{p})\!\max\!\left\{ \!\gamma(k,l)\!-\!1,0\right\} \label{eq:xi_decision_directed}\end{equation}
using the modifications proposed in \cite{CohenNonStat2001} to take
into account speech presence uncertainty.

\subsection{Optimal gain modification under speech presence uncertainty}

\label{sec:Optimal-gain-modification}In order to take into account
the probability of speech presence, we derive the estimator for the
loudness domain:\emph{\begin{equation}
\hat{A}(k)=\left(E\left[A^{\alpha}(k)\left|Y(k)\right.\right]\right)^{\frac{1}{\alpha}}\label{eq:optimal-gain}\end{equation}
}

Considering $H_{1}$, the hypothesis of speech presence for source
$m$, and $H_{0}$, the hypothesis of speech absence, we obtain:\begin{eqnarray}
E\left[\left.\! A^{\alpha}(k)\right|\! Y(k)\right] & = & p(k)E\left[\left.A^{\alpha}(k)\right|H_{1},Y(k)\right]\nonumber \\
 & + & \left[1-p(k)\right]\! E\!\left[\left.\! A^{\alpha}(k)\right|\! H_{0},\! Y(k)\right]\label{eq:optimal-gain2}\end{eqnarray}
where $p(k)$ is the probability of speech at frequency $k$.

The optimally modified gain is thus given by:\begin{equation}
G(k)=\left[p(k)G_{H_{1}}^{\alpha}(k)+(1-p(k))G_{min}^{\alpha}\right]^{\frac{1}{\alpha}}\label{eq:optimal_gain3}\end{equation}
where $G_{H_{1}}(k)$ is defined in Eq. \ref{eq:Gain_H1}, and $G_{min}$
is the minimum gain allowed when speech is absent. Unlike the log-amplitude
case it is possible to set $G_{min}=0$ without running into problems.
For $\alpha=1/2$, this leads to:\begin{equation}
G(k)=p^{2}(k)G_{H_{1}}(k)\label{eq:optimal_gain_final}\end{equation}

Setting $G_{min}=0$ means that there is no arbitrary limit on attenuation.
Therefore, when the signal is certain to be non-speech, the gain can
tend toward zero. This is especially important when the interference
is also speech since, unlike stationary noise, residual babble noise
always results in musical noise.

The probability of speech presence is computed as:\begin{equation}
p(k)=\left\{ 1+\frac{\hat{q}(k)}{1-\hat{q}(k)}\left(1+\xi(k)\right)\exp\left(-\upsilon(k)\right)\right\} ^{-1}\label{eq:def_speech_prob}\end{equation}
where $\hat{q}(k)$ is the \emph{a priori} probability of speech presence
for frequency $k$ and is defined as:\begin{equation}
\hat{q}(k)=1-P_{local}(k)P_{global}(k)P_{frame}\label{eq:def_q_a_priori}\end{equation}
where $P_{local}(k)$, $P_{global}(k)$ and $P_{frame}$ are defined
in \cite{CohenNonStat2001} and correspond respectively to a speech
measurement on the current frame for a local frequency window, a larger
frequency and for the whole frame.

\section{Results}

\label{sec:Results}The system is evaluated in a context of mobile
robotics, where an array of 8 microphones is mounted on a mobile robot.
In order to test the system, 3 voices (2~female, 1~male) were recorded
separately, in a quiet environment. The background noise was recorded
on a mobile robot and is comprised of room ventilation and internal
robot fans. All four signals were recorded using the same microphone
array and subsequently mixed together to allow SNR and distance mesures.

In evaluating our post-filter, we use both the segmental SNR and the
log spectral distortion (LSD), which is defined as:\begin{equation}
LSD=\frac{1}{L}\sum_{l=0}^{L-1}\left[\frac{1}{K}\sum_{k=0}^{K-1}\left(20\log_{10}\frac{\left|X(k,l)\right|+\epsilon}{\left|\hat{X}(k,l)\right|+\epsilon}\right)^{2}\right]^{\frac{1}{2}}\label{eq:def_LSD}\end{equation}
where $L$ is the number of frames, $K$ is the number of frequency
bins and $\epsilon$ is meant to prevent extreme values.

\begin{table}

\caption{Log spectral distortion and segmental SNR for each of the 3 separated
sources.\label{cap:Log-spectral-distortion}}

\begin{tabular}{|c|c|c|c|}
\hline 
LSD/SegSNR (dB)&
female 1&
female 2&
male 1\tabularnewline
\hline
\hline 
Mic. input&
23.4/-5&
21.6/-5&
21.6/-6.2\tabularnewline
\hline 
LSS output&
19.2/2.5&
17.1/4.1&
17.5/1.6\tabularnewline
\hline 
1-ch. post-filter&
10.4/6.1&
9.4/7.7&
9.9/4.1\tabularnewline
\hline 
Cohen p-f&
8.9/6.4&
9.7/4.7&
10.3/4.5\tabularnewline
\hline 
Proposed p-f&
6.5/7.6&
6.7/8.1&
7/7.1\tabularnewline
\hline
\end{tabular}
\end{table}

\begin{figure*}[t]
a)\includegraphics[%
  width=0.40\paperwidth,
  height=0.14\paperwidth]{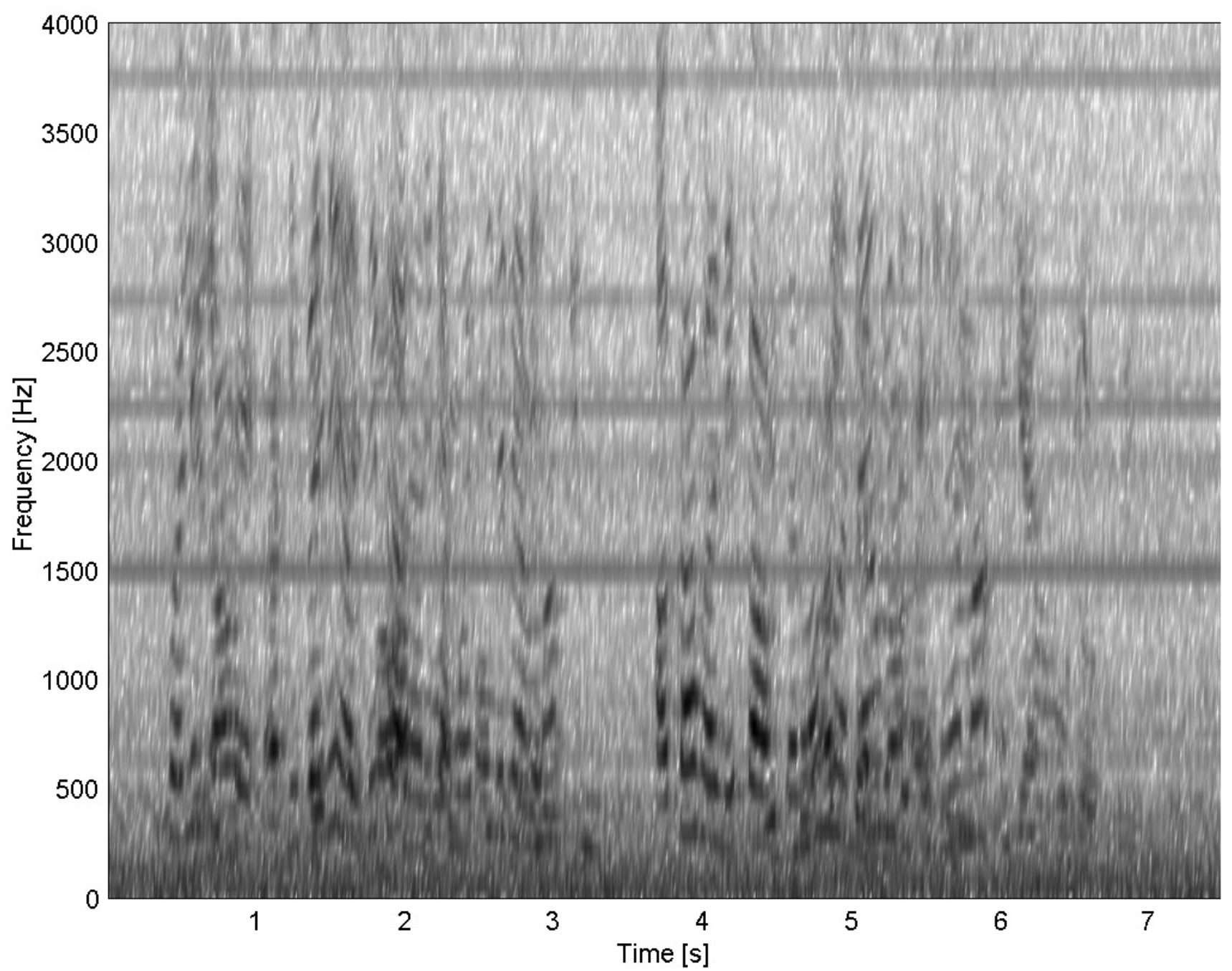}d)\includegraphics[%
  width=0.40\paperwidth,
  height=0.14\paperwidth]{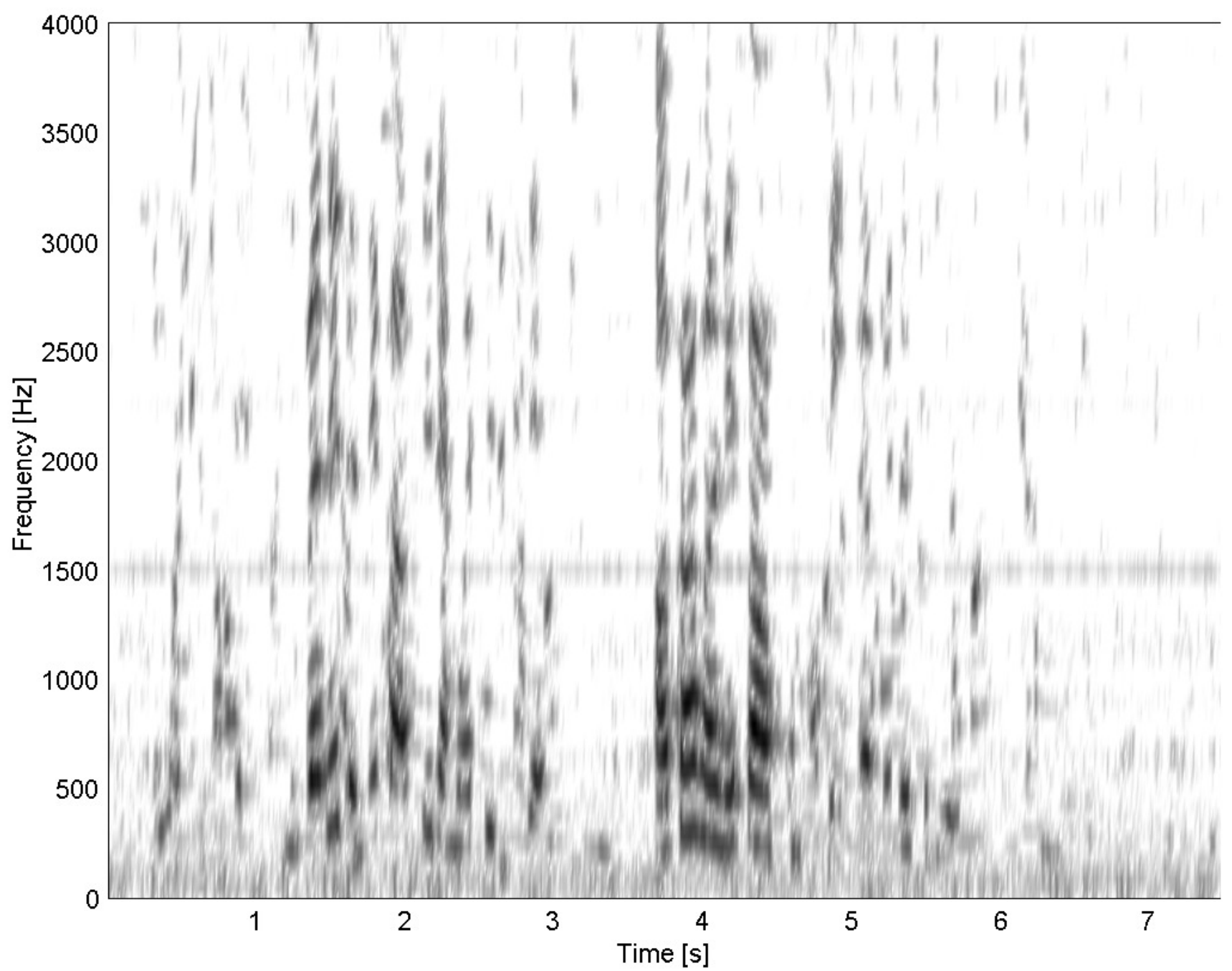}

b)\includegraphics[%
  width=0.40\paperwidth,
  height=0.14\paperwidth]{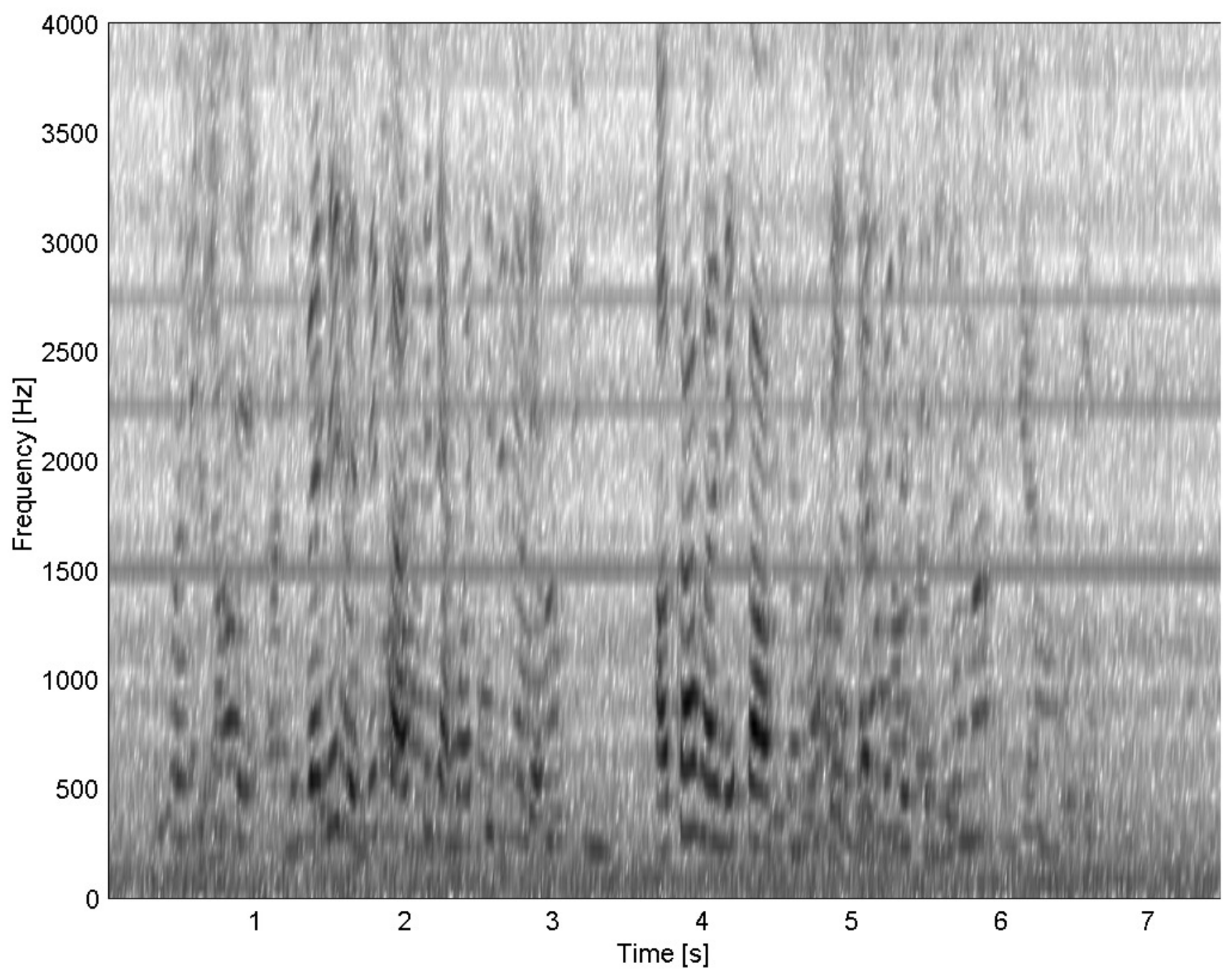}e)\includegraphics[%
  width=0.40\paperwidth,
  height=0.14\paperwidth]{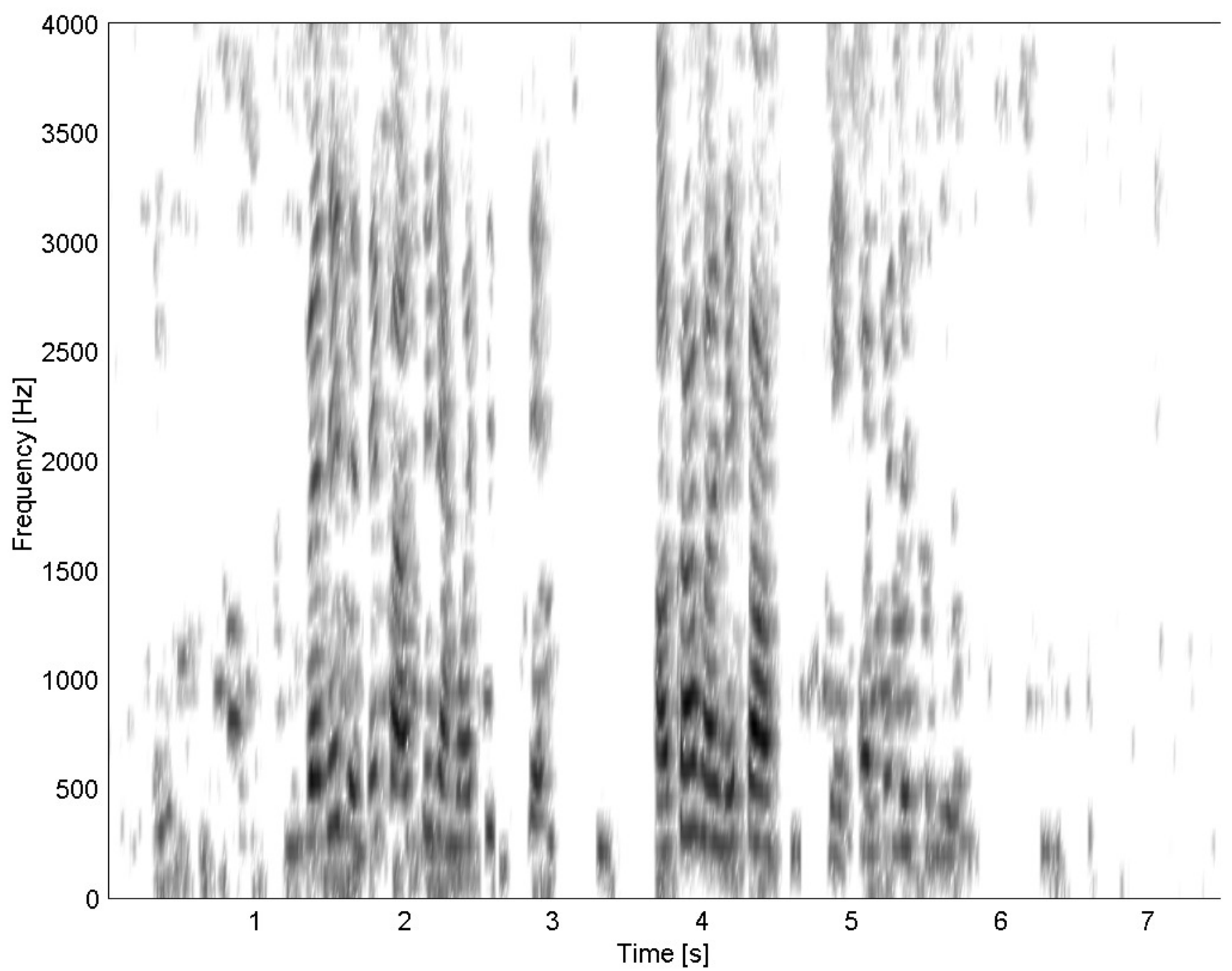}

c)\includegraphics[%
  width=0.40\paperwidth,
  height=0.14\paperwidth]{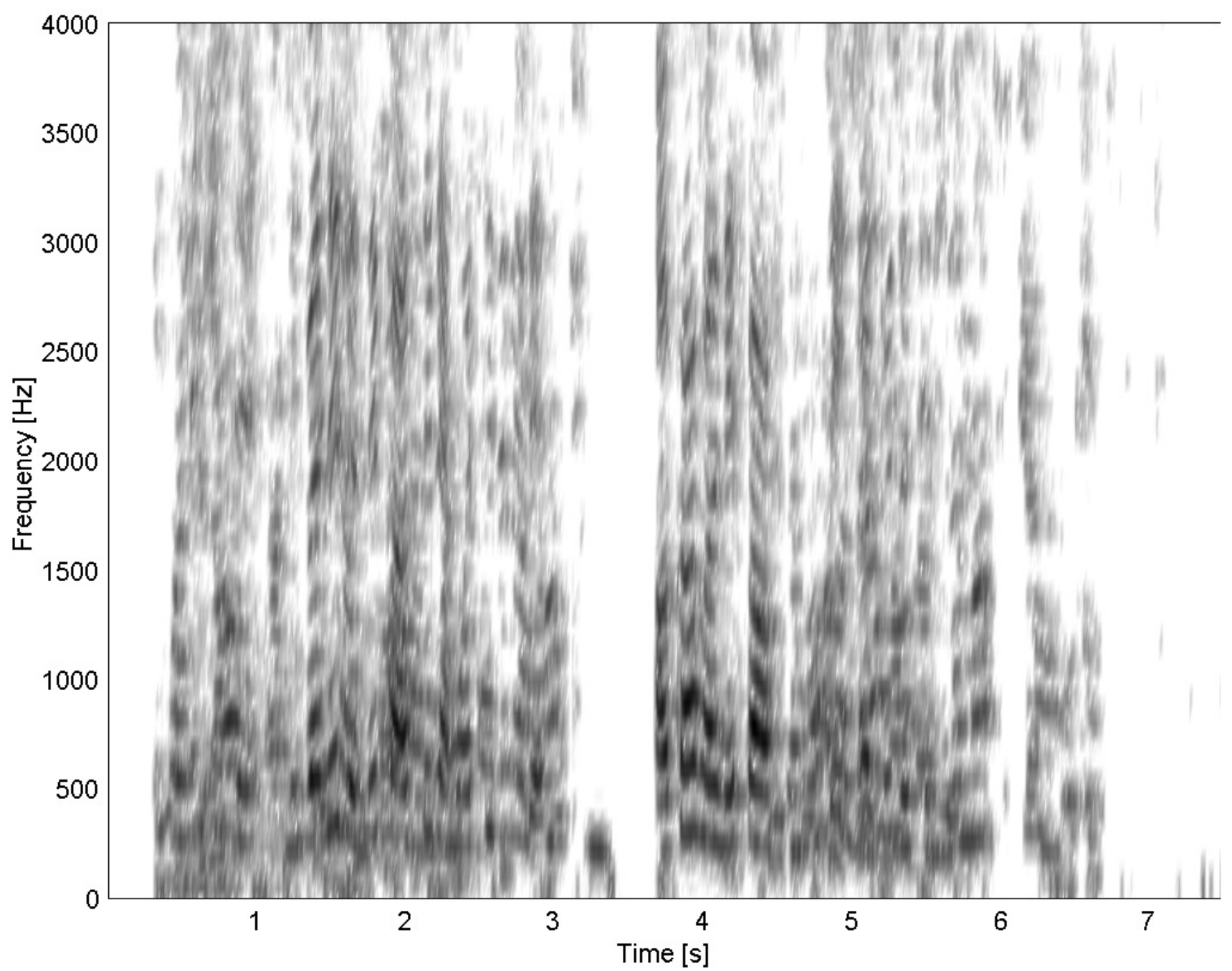}f)\includegraphics[%
  width=0.40\paperwidth,
  height=0.14\paperwidth]{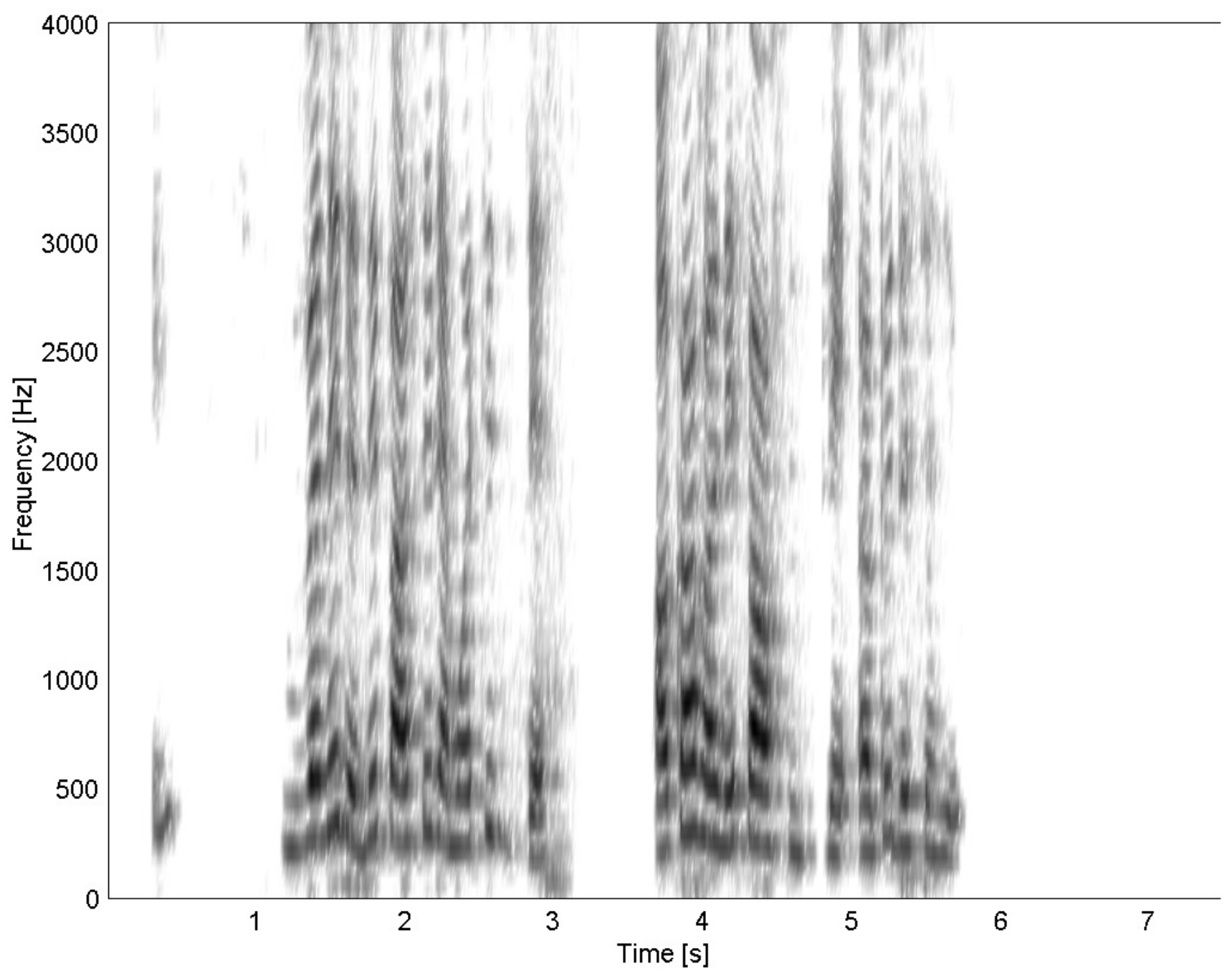}

\caption{Spectrogram for separation of first source (female voice) (a) Average
of microphone inputs (b) Linear separation output (c) Single-channel
post-filtering (d) Adaptation of Cohen post-filter (e) Proposed post-filter
(f) Reference signal\label{cap:Spectrogram-for-separation}}
\end{figure*}

Table \ref{cap:Log-spectral-distortion} compares the results for
separation of each of the 3 original sources with the single-channel
and the multi-channel Cohen post-filters, both described in \cite{CohenArray2002}.
The Cohen post-filter is adapted to uses the other sources as reference
noise signals. The improvement of our post-filter in terms of LSD
and SegSNR are confirmed by informal listening. 

The spectrograms for the first source (female) is shown in Figure
\ref{cap:Spectrogram-for-separation}. Even though the task involves
non-stationary interference with the same frequency content as the
signal of interest, we observe that our method (unlike the single-channel
post-filter) is able to remove most of the interference, while not
causing excessive distortion to the signal of interest. Also, for
this task, we explain the improvement of our post-filter over the
Cohen multi-channel post-filter by the fact that the interference
is adaptively estimated even in the presence of the source of interest.
This is not the case with the Cohen post-filter, for which the noise
estimator (for both stationary and transient noise) is only adapted
when the source of interest is absent.

\section{Conclusion}

\label{sec:Discussion}We proposed a microphone array post-filter
designed in the context of separation of multiple simultaneous sources.
It is based on a loudness-domain MMSE estimator in the frequency domain
with a noise estimate that is computed as the sum of a stationary
noise estimate and an estimation of leakage due to the linear source
separation (LSS) algorithm. Experimental results show a reduction
in log spectral distortion of up to $12\:\mathrm{dB}$ compared to
the output of the LSS and up to $4\:\mathrm{dB}$ over the single-channel
post-filter.

The proposed post-filter is general enough to be applicable to most
source separation algorithms. A possible improvement to the algorithm
would be to derive a method that automatically adapts the leakage
factor $\eta$ to track the leakage of an adaptive LSS algorithm.

\bibliographystyle{IEEEbib}
\bibliography{icassp}

\end{document}